\RequirePackage{fix-cm}
\documentclass[smallextended]{svjour3}       % onecolumn (second format)
\smartqed  % flush right qed marks, e.g. at end of proof
\usepackage{graphicx}
\begin{document}

\title{A Mission to Nature's Telescope for High-Resolution Imaging of an Exoplanet
}
%\subtitle{Imaging an Exoplanet}


\author{Louis D. Friedman       \and
        Darren Garber \and
        Slava G. Turyshev \and
        Henry Helvajian \and
        Thomas Heinshiemer \and
        John McVey \and
        Artur R. Davoyan
}


\institute{Louis D. Friedman
           \at
             The Planetary Society (emeritus), Pasadena, 91101, CA USA.\\
           %  Tel.: +123-45-678910\\
           %  Fax: +123-45-678910\\
%              \email{louisdfriedman@gmail.com}           %  \\
           \and
            Darren Garber
           \at Xplore Inc, Mercer Island, 98040, WA USA.
           %  Tel.: +123-45-678910\\
           %  Fax: +123-45-678910\\
%              \email{darren.garber@xplore.com}   
            \and
            Slava G. Turyshev
           \at Jet Propulsion Laboratory, California Institute of Technology, Pasadena, 91109, CA USA.
           %  Tel.: +123-45-678910\\
           %  Fax: +123-45-678910\\
%              \email{turyshev@jpl.nasa.gov}   
            \and
            Henry Helvajian, Thomas Heinshiemer, John McVey 
           \at The Aerospace Corporation, El Segundo, 90245, CA USA.
           %  Tel.: +123-45-678910\\
           %  Fax: +123-45-678910\\
%              \email{henry.helvajian@aero.org, thomas.heinsheimer@aero.org, john.p.mcvey@aero.org}                 
            \and
            Artur R. Davoyan
           \at Department of Mechanical and Aerospace Engineering, University of California, Los Angeles, 90095, CA USA.
           %  Tel.: +123-45-678910\\
           %  Fax: +123-45-678910\\
%              \email{davoyan@seas.ucla.edu}  
}

\date{Received: date / Accepted: date}

\maketitle

\begin{abstract}
The solar gravitational lens (SGL) provides a factor of $10^{11}$ amplification for viewing distant point sources beyond our solar system.  As  such, it may be used for resolved imaging of extended sources, such as exoplanets, not possible otherwise. To use the SGL, a spacecraft carrying a modest telescope and a coronagraph must reach the SGLs focal region, that begins at $\sim$550 astronomical units (AU) from the Sun and is oriented outward along the line connecting the distant object and the Sun.  No spacecraft has ever reached even a half of that distance; and to do so within a reasonable mission lifetime (e.g., less than 25 years) and affordable cost requires a new type of mission design, using solar sails and microsats ($<100$~kg). The payoff is high -- using the SGL is the only practical way we can ever get a high-resolution, multi-pixel image of an Earth-like exoplanet, one that we identify as potentially habitable.  This paper describes a novel mission design starting with a rideshare launch from the Earth, spiraling in toward the Sun, and then flying around it to achieve solar system exit speeds of over $20$ AU/year.  A new sailcraft design is used to make possible high area to mass ratio for the sailcraft.  The mission design enables other fast solar system missions, starting with a proposed very low cost technology demonstration mission (TDM) to prove the functionality and operation of the microsat-solar sail design and then, building on the TDM, missions to explore distant regions of the solar system, and those to study  Kuiper Belt objects (KBOs) and the recently discovered interstellar objects (ISOs) are also possible. 

\end{abstract}

\section{Introduction}
\label{intro}

Going to a potentially habitable exoplanet is impractical even when the target exoplanet is one of the closest ones. More likely an identified potentially habitable Earth-like planet would be ten to a hundred times further away than the nearest star.   Nor can we, for any reasonable cost, build a telescope large enough to resolve features on an exoplanet. The largest optical telescope on Earth is the Large Synoptic Survey Telescope (LSST) in Chile\footnote{LSST is the Vera C. Rubin Observatory, https://www.lsst.org/} -- with an 8.4-meter aperture.  The resolving power of a telescope is given by the ratio $ \simeq \lambda/d$, where $\lambda$ is the wavelength of the light and $d$ is the  telescope diameter.  For the shortest visible wavelength (near UV) that is $\lambda=400$ nanometers (nm).  Dividing that by 8.4 meters gives an angular resolution of $\lambda/d=47.6$ nanoradian (nrad). Multiplying that by the distance to the nearest exoplanet (i.e., Proxima Centauri b at 40 quadrillion meters from us), one gets a linear resolution at that distance of 1.9 million km, $\sim 150$ times the size of the Earth.  Even with the largest telescopes that are being built on the ground -- the European Extremely Large Telescope  (ELT)\footnote{European Extremely Large Telescope  (ELT), first light in 2025, https://elt.eso.org/} with its primary mirror of 39.3~m and the Thirty Meter Telescope (TMT)\footnote{Thirty Meter Telescope (TMT), first light in 2027, https://www.tmt.org/} the situation will improve only by a factor of $\sim3$. That is, the world's largest telescopes, even theoretically, would not be able to resolve a whole Earth-sized planet in our stellar neighborhood.  And of course, these telescopes on the ground are much bigger than anything we could put in space.  As a result, we are many orders of magnitude away from being able to resolve or visually see any surface details on an exoplanet.  

Fortunately, however, nature provides a solution.  There is a ``natural'' telescope in our solar system that provides light amplification of over 100 billion power.  All we have to do is get there so we can use it. 

\begin{figure}[t]
	\centering
	        \includegraphics[scale=0.35]{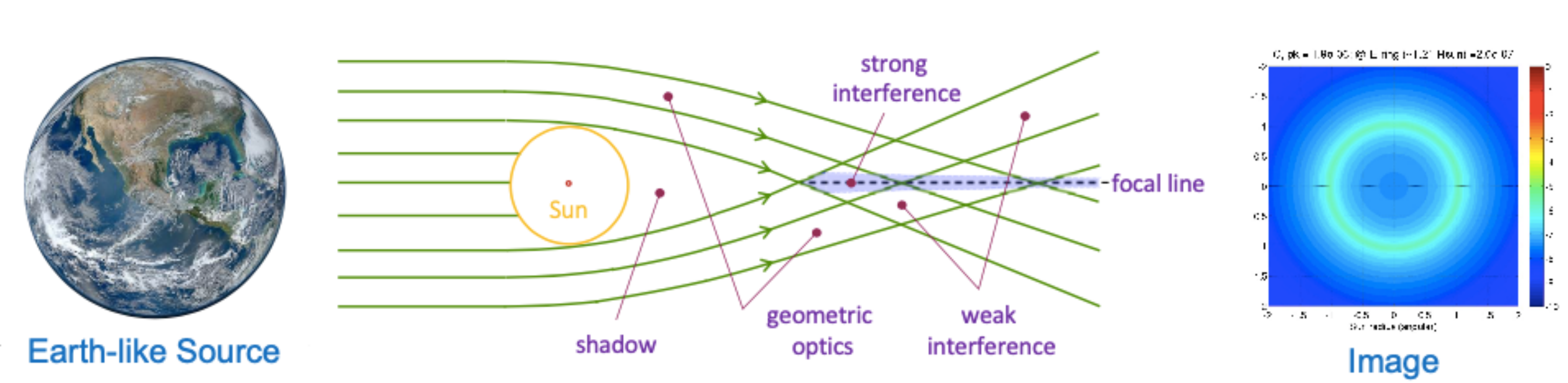}
	  \caption{The geometry of the solar gravity lens used to form an image of a distant object in the Einstein ring.}
	  \label{fig1}
\end{figure}

When Albert Einstein published his theory of general relativity in 1916 \cite{Einstein-1916}, he included several predictions from the theory which could be tested by accurate scientific measurements.  One of those predictions asserts that light rays, which are made up of zero mass photons, would deviate from the Newtonian straight line on which they were thought to travel, onto an Einsteinian curved line -- bent by the gravitational field in which they are traveling.  Close to a large object, like the Sun, the light ray would be bent by an amount that could be measured.  Three years later, the British astronomer, Sir Arthur Eddington measured that bending, exactly as predicted by the theory \cite{Eddington:1919}.  The bending of light rays as they pass close to the Sun means that the Sun acts like a lens, curving the rays by a small angle $\theta=4GM_\odot/c^2b$, where $G$ is the gravitational constant, $c$ is the speed of light,  $M_\odot$, is the mass of the Sun, and $b$ is the light ray's impact parameter (defined as the perpendicular distance between the path of a photon at emission and the center of a the Sun). This bending results in focusing the rays behind the Sun at a distance of $r\simeq b/\theta=b^2c^2/4GM_\odot$ \cite{Turyshev-Toth:2017}. Light rays from a distant object (light years away) will pass by the limb at varying distances, so the focus is not just a point, but a series of points which results in a focal line extending indefinitely outward away from the Sun limited by spatial extend f the object, see Fig.~\ref{fig1}.  Plugging in the numbers and defining $R_\odot$ to be the solar radius, the calculation results in the focal line beginning at $r=547 ~(b/R_\odot)^2$ astronomical units (AU) from the Sun and extending outward on the straight line formed by the distant object and the Sun \cite{Turyshev-Toth:2019-fin-difract}. The light from a point source  is amplified at that focus \cite{Turyshev-Toth:2017} by a factor of $8\pi^2 GM_\odot/ c^2\lambda= 2.92\times 10^{11} \,(400\,{\rm nm}/\lambda)$, nearly 300 billion! The resolving power of the SGL is given by $\sim\lambda/b=0.04 \, (\lambda/400\,{\rm nm}) (R_\odot/b)$~nrad \cite{Turyshev-Toth:2020-extend}, which may not be achieved by any technological means today.   
 
 Relying on the SGL, an extended object could be resolved to a scale of few kilometers on its surface \cite{Toth-Turyshev:2020}.  If that distant object is an exoplanet, that would mean one could see regional features, the size of continents, seas, forests, or even cities, on an exoplanet -- not just the nearest exoplanet, but on thousands of them within the radius of tens of light-years from Earth.  So, all one has to do to get high-resolution images of potentially habitable exoplanets is send spacecraft with a modest telescopes (1--2 meters) out that far to the desired focal line.  
 
 This paper discusses the imaging scenario and the design of a mission that  could provide high-resolution imaging of a distant habitable exoplanet. The paper is organized as follows:  In Section~\ref{sec:obs-scenario} we discus mission design considerations and present a science observing scenario, characterize the background environment and consider various propulsion options. In Section~\ref{sec:sailcraft} we address the questions that drive the sailcraft design and operations. In Section~\ref{sec:concl}, we conclude and discuss the near future efforts  toward the mission to the SGL that includes several demonstrations flights to various destinations in the solar system including flybys of the Kuiper Belt objects (KBOs) and studies of the recently discovered interstellar objects (ISOs).

\section{Mission Design Considerations}
\label{sec:obs-scenario}

\subsection{Science observing scenario}
\label{sec:sci-obs-scen}

\begin{figure}%[<options>]
	\centering
	        \includegraphics[scale=0.40]{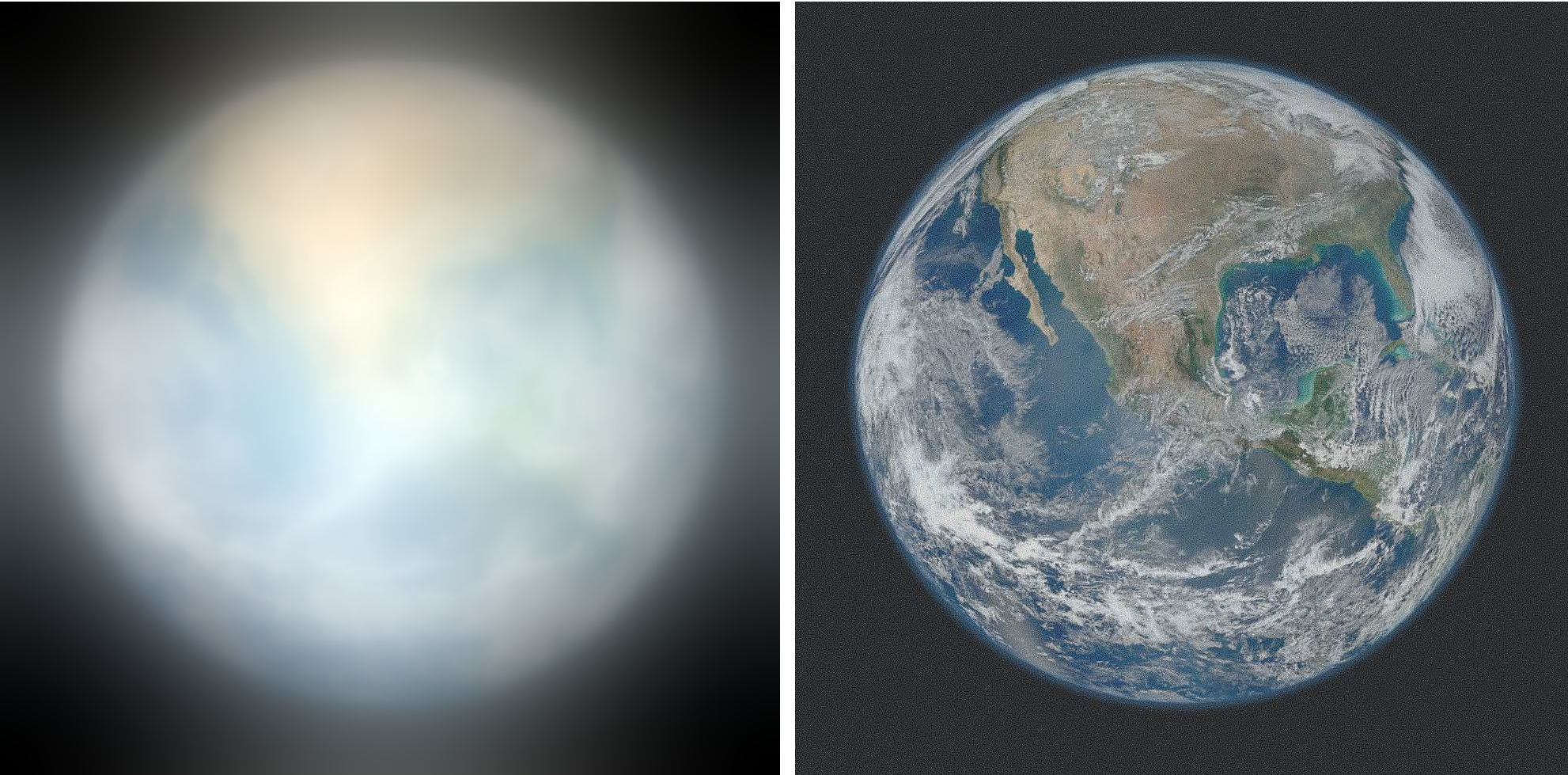}
	  \caption{Left: Simulated image of an Earth-like exoplanet at 650 AU, as would be seen using the SGL. Right: Recovered image of the Earth after deconvolution. (Adapted from \cite{Turyshev-Toth:2020-extend,Toth-Turyshev:2020}.)}
	  \label{fig2}
\end{figure}

To utilize the Sun as an optical instrument, light must be collected on the opposite side of the Sun relative to the preselected exoplanetary target. As we discussed above, the SGL's focal region forms a half-line that begins at $\sim$547 AU from the Sun and extends well beyond 1500 AU.  
Figure~\ref{fig2} shows the light rays (light from the parent star reflected by the exoplanet) of the Earth-like planet being bent as they pass by the Sun to form the image around the focal line emanating outward from 547 AU.

A mission to the SGL focal region, that we will call here -- the SGLF mission \cite{Turyshev-etal:2020-PhaseII}, would begin with the selection of the intended target, an Earth-like exoplanet orbiting within the habitable zone of its host-star and likely showing the presence of an atmosphere or other signs of life-bearing conditions. The SGLF mission would then be launched in the appropriate direction. During its estimated $\sim$25-year cruise (at speeds of $\geq20$~AU/yr), the parent star's location would be observed with 1 microarcsecond ($\mu$as) precision at least 100 times, so that its position would be known at the 0.1 $\mu$as level, equivalent to $\sim$45 km at 30 parsec (pc), which is within the present capabilities. The orbital period of the planet would be known to an accuracy of $<$1--3\%.

The envisioned SGLF spacecraft will carry a telescope with a meter-class aperture, large enough to resolve the solar disk from the SGLF location, so that its light can be blocked by a suitable coronagraph. The instrument will observe light from the intended target in the form of a faint Einstein-ring surrounding the Sun.  There are a few important elements to consider: 
\begin{itemize}
\item At a particular heliocentric distance, $z$, at the focal region of the SGL, an image of an Earth-like exoplanet positioned at distance $z_0$ from the Sun is compressed to a cylinder with a diameter of $d_{\rm im}=2R_\oplus (z/z_0)$, which for $z=650$~AU and $z_0=100$~light years results in $d_{\rm im}\simeq 1.3$\,km, which, of course, is much larger than a meter-class telescope of the SGLF. 
\item The SGLF spacecraft must be positioned within the image cylinder and, while continue to move along the focal line, it also moves within the image in a pixel-by-pixel by pixel fashion. The objective is to collect the brightness data at each pixel, which is the primary data that will be used to produce a high-resolution image of the entire exoplanet \cite{Toth-Turyshev:2020,Turyshev-Toth:2020-image,Turyshev-Toth:2020-extend}. 
\item 
 The imaging done on a pixel-by-pixel basis, where the brightness is received from a thin annulus circumventing the Sun, known as the Einstein ring. The diameter of the ring is slightly larger than that of the Sun, the thickness of the annulus is the same as the image diameter, $d_{\rm im}$, which, of course, depends on the target exoplanet and may be up to few kilometers.
\item The Sun's limb isn't sharp, it is softened due to the solar corona. So, when the SGLF telescope observes the Einstein ring around the Sun, the light from the ring arrives at the telescope together with light from the solar corona. Thus, the brightness of the corona becomes a source of noise. Luckily, by positioning the meter-class SGLF telescope at larger heliocentric distances, one can observe the Einstein ring at larger separations from the Sun, where the solar corona noise is reduced. This consideration pushes the beginning of the practical observing region for the SGLF mission beyond 547 AU, close to 650 AU \cite{Turyshev-Toth:2019,Turyshev-Toth:2020-extend};
\item The spacecraft must not just reach the focal line, it must then fly along the focal line of the SGL while looking at the ring as it extends outward, collecting brightness data with the image pixels over a period of weeks, months or even years;
\item And, of course, the spacecraft has to get to 650~AU and beyond in a reasonable amount of time -- say the working lifetime of the mission scientists, e.g., 25-40 years.  That works out to an average speed on 20--30 AU/year.  The fastest spacecraft we have ever sent out of the solar system is the Voyager spacecraft, traveling at about 3.5 AU/year.  
Accomplishing a speed seven times our current best record ($\sim$418,000 km/hr) is a difficult feat to achieve, but much less than the interstellar speed the interstellar speed of about 13,000 AU/year that would be required to reach the nearest star in 20 years. High-resolution images would be possible continuously, analogous to that of a planetary orbiter mission.
\end{itemize}

The imaging is more complicated than snapping pictures with a camera, as we now routinely do while visiting the planets in our solar system.  The characteristics of gravitational light bending spreads in the image formed around the focal line. Consequently, each pixel from the whole exoplanet is similarly distributed in the image resulting in image blurring. The collected data forms a blurred  image which will be deconvolved \cite{Turyshev-Toth:2020-extend} (Fig.~\ref{fig2}).  

Consider an Earth-like exoplanet $d_{\rm p}\simeq12,756$ km diameter planet placed at the distance of $r_{\rm p}=100$ light years away from us and observed at a heliocentric distance of $r_{\rm obs}=650$ AU from the focal region of the SGL, the diameter of the image is $d_{\rm im}=(r_{\rm obs}/r_{\rm p}) d_{\rm p}\sim 1.31 \,(r_{\rm obs}/650~{\rm AU})$~km. Clearly, as we move outward, $r_{\rm obs}$ increases, and thus the image become slightly larger. For instance, by the time we get to 900 AU it is already $\sim1.82$~km.  

That image is convoluted with the SGL's point-spread function (PSF) that causes blurring, which and must be deconvolved by image processing algorithms \cite{Toth-Turyshev:2020} -- an integration of data that takes place over many months or even years as it is accumulated from the rotating planet (see details in \cite{Turyshev-Toth:2020-extend}).  At the beginning the image is blurred and looks like the picture on the left of Fig.~\ref{fig2}, but with appropriate integration times, data with sufficient quality is accumulated and can be used for deconvolution, which will be used  to recover a true image, like that on the right of Fig.~\ref{fig2} will show the features of the Earth-like planet.  

It still takes a telescope to get this image -- but one that fits on a small spacecraft -- with the diameter of 1--2 meters. In addition, a coronagraph is needed -- that is, the Sun, which of course is in the center of the telescope's field of view, has to be blocked out.  That could be done either with an external coronagraph, called a star shade -- but that would require a second spacecraft perfectly positioned in front of the telescope.  A more appropriate option is an internal coronagraph \cite{Turyshev-etal:2020-PhaseI}. A coronagraph blocks out the central rays of light internally within the telescope by computing the disk's location.  It won't be perfect, because the Sun has a dynamically varying corona -- but with sufficient integration time that can be modeled as well.  

\subsection{Relevant Dynamics and Environment}

Even with the relatively small size of the Einstein ring, the spacecraft and telescope must be maneuvered over the distance of tens of kilometers  to collect pixel-by-pixel all the data necessary to construct the image. This is needed as the image moves because of the multiple motions are present, namely 1) the planet orbits its parent star, 2) the star moves with the respect to the Sun, and 3) the Sun itself is not static, but moves with respect to the solar system barycentric coordinates. To compensate for these motions, the spacecraft will need micro-thrusters and electric propulsion, the solar sail obviously being useless for propulsion so far from the Sun.  The sail may be jettisoned long before arrival in the focal region, although we are studying ways to repurpose it as an antenna for radio communication or a reflector for laser communication. Power for electric propulsion would be provided by the same radioisotope source (radioisotope thermoelectric generators (RTG) or radioisotope heater units (RHU)) that provides power to the spacecraft communications, instruments, and computer.

 \begin{figure}%[<options>]
	\centering
	        \includegraphics[scale=0.70]{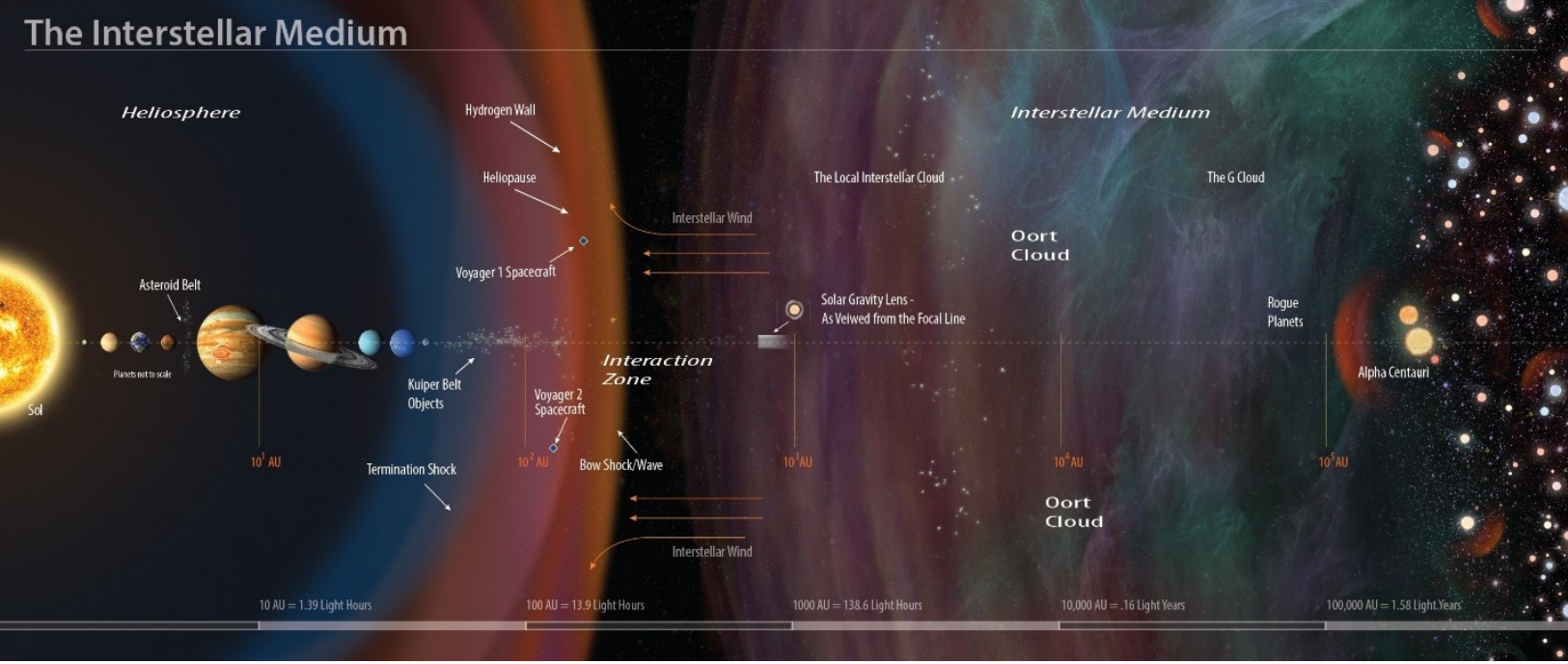}
	  \caption{Artist's depiction of the solar system through the interstellar medium (ISM) to the nearest star (logarithmic scale).}
	  \label{fig3}
\end{figure}

As mentioned, the solar gravity lens focus is located in the interstellar medium beginning at 547 AU.  The location is depicted in Fig.~\ref{fig3}.  
Note that the scale in that figure is logarithmic -- otherwise the distances in the solar system wouldn't show up on a map depicting distances in light-years. On this scale the SGL focal region appears roughly halfway to the nearest star. Even more noticeable is that there is not much else on the way.  The interstellar medium is almost empty -- particles from the Sun forming the solar wind to about 120 AU where they meet particles from other stars (forming the interstellar wind).  The boundary, a shock wave where those two streams collide is called the heliopause.  The interstellar medium itself is an important area for scientific study.  Heliophysicists want to understand the composition and the dynamics of the stellar wind that blows here from other stars in our galaxy.  Only the Voyager spacecraft have dipped a little way into the interstellar medium, the scientists would like to get two to three times further to sample that stellar wind separate from the solar wind of our own solar system.  
Some define that heliopause as the boundary to interstellar space, although gravitationally the Sun remains dominant a lot further, beyond $10^5$ AU. 
 
 \subsection{Propulsion considerations}
 
Interstellar travel is a bridge too far, but its lure is compelling.  Just as compelling is the goal of the discovery and investigation of life on other worlds.  To see such life, or at least evidence about it, the solar gravity lens is required.  A mission there might be the next step once a likely life-bearing planet is identified. It is a challenging mission reaching beyond where no one has gone before (or even yet proposed to) but still thousands of times closer and less challenging than the interstellar flight goal.  It can also be targeted to the primary exoplanet of interest, not just the nearest one. Getting this far this fast requires either some big thinking, or alternatively to ``big,'' clever and nimble thinking.  

The big way is with a powerful, large launch vehicle and a spacecraft propulsion system capable of supplying a lot of thrust for the long voyage.  A chemicals propulsion system (as most spacecraft have) would use a solid or liquid propellant motor to supply the $\Delta v$, the change in velocity necessary in deep space.  The most efficient place to apply the $\Delta v$ is when the spacecraft is closest to the Sun, at its perihelion. Enough $\Delta v$ there can open the elliptical orbit to make it hyperbolic extending out of the solar system.\footnote{Modern the chemical rocket cannot produce enough $\Delta v$ for realistic spacecraft mass to go fast enough to reach the SGL focal region in $\sim$30 years. These powerful rocket engines, even many of them staged together, may produce $\Delta v$ on the order of 13 AU/year at best.}
But a big rocket motor working near the Sun will need protection from the heat of the Sun (as will the spacecraft itself) which means a many hundred-kilogram heat shield must be carried -- a big weight that requires even a bigger motor and a bigger (and more expensive) launch vehicle.  

Electric propulsion, using plasma jets could be used but they too would either need to go close to the Sun for the $\Delta v$ or operate for a long time deep in space where solar power would not be possible and therefore nuclear power would be required.  Powerful nuclear rockets are also heavy and relatively expensive.  Currently, neither nuclear electric nor nuclear thermal propulsion (both of which require nuclear reactors) are being used in space.  They are not prohibited, but they come with environmental and safety concerns. The USA, Russia, and China are developing the capability for future nuclear propulsion missions, but the time scale is somewhat uncertain, the weight requirements for launch are large, and as noted, it may be expensive.  NASA's Voyager and New Horizons missions used RTGs, a form of nuclear power not requiring a reactor, but just using the decay heat from radioactive material in the heat source.  That supplies enough power for spacecraft operations, but not enough for the propulsion to reach the speeds necessary for solar system escape.  
 
 \begin{figure}%[<options>]
	\centering
	        \includegraphics[scale=1.50]{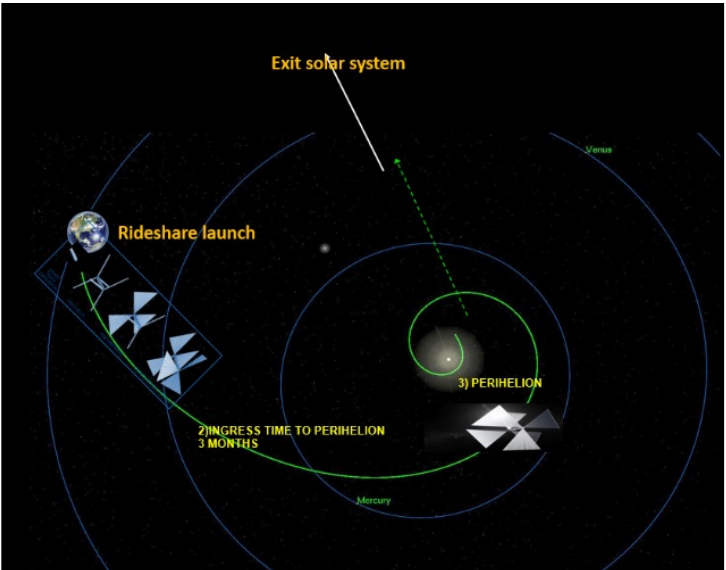}
	  \caption{Trajectory of the mission design concept for a solar sailcraft to exit the solar system.}
	  \label{fig4}
\end{figure}

An alternative is to think small -- use the newly developed microsat and CubeSat capability for spacecraft (with weight less than 100 kg) and a solar sail with large enough area for putting the microsat onto a hyperbolic trajectory leaving the solar system.  The solar sail would also need to apply most of its $\Delta v$ close to the Sun, but nowhere near as close as would be required for the chemical propulsion rocket motor \cite{Friedman-Garber:2014}. Not only does it benefit from the dynamics of applying the $\Delta v$ at perihelion, but the solar power is much greater close to the Sun and applies more force to the trajectory.  

A microsat -- solar sail mission can reach high solar system exit velocity and get to the solar gravity lens focus with modest cost, and without requiring heavy launch vehicles or nuclear propulsion.  The mission concept is depicted in Fig.~\ref{fig4};  the mission launches on its heliocentric trajectory with minimum launch energy (denoted by rideshare\footnote{Rideshare is a term used when a small payload hitches a ride with a bigger payload thereby saving launch costs.} in the figure). The sail tacks inward toward the Sun until it reaches is closest approach (perihelion).  At that point the sail is reoriented to pick up maximum solar pressure and increases the velocity.  The amount of the increase is proportional to the sail area, $A$, divided by the spacecraft mass, $m$, and is larger when the perihelion distance is smaller.  

The SGLF mission is the most demanding because we want to go far out fast.  But a whole class of fast solar system missions with lesser requirements are possible with the same mission design.  The timing of the trajectory is flexible -- one could drop into solar orbit or make out of plane maneuvers while on the ingress trajectory.   The speed at which it flies is 
$A/m$, the sail should be as large as possible and the spacecraft mass as small as possible.  The sail should also be thin and lightweight so that it won't take up much of the payload mass.  The force comes from the transfer of momentum from the sunlight photons as they bounce off the highly reflective sail.  Just like on a terrestrial sailboat the sail can be pointed so that the ship (spacecraft) can tack either toward the Sun (inward in the solar system) or away (outward).  Except here it is not the wind, but sunlight pushing the sail. (The solar wind is a three orders of magnitude smaller force coming from electrons and protons from the Sun.)  A sailboat operates at the interface of wind and water, with its rudder setting the direction.  

\section{The SailCraft}
\label{sec:sailcraft}

The solar sailcraft operates at the interface of sunlight and gravity, the latter resulting in orbital motion around the Sun.  If the sail is tilted so that the force adds to the orbit velocity the spacecraft goes outward (the orbit gets bigger).  If it subtracts from the orbit velocity the spacecraft falls inward in a smaller orbit.  That is how the spacecraft can be steered and how it can be made to fall inward from the Earth to the Sun, and then outward after its closest passage to the Sun.  As mentioned earlier if the force is big enough the orbit gets so big that it opens out into a hyperbola on which the spacecraft can fly indefinitely out of the solar system.  The bigger the sail, the smaller the mass and the closer we can get to the Sun make the force larger and sufficient for solar system escape.  The bigger that force the faster the spacecraft will go on its exit hyperbola.  Once on the exit trajectory moving away from the Sun the force drops fast, by the square of the distance, so that once past Mars there is very little force due to the solar radiation pressure as it decreases being inversely proportion to the square of the heliocentric distance.

The bigger the sail, the lighter the spacecraft, and the closer we can get it to the Sun -- the higher the velocity and the faster we go.  This is the spacecraft design problem.  

\subsection{How large can we make the sail?}

\begin{figure}%[<options>]
	\centering
	        \includegraphics[scale=0.60]{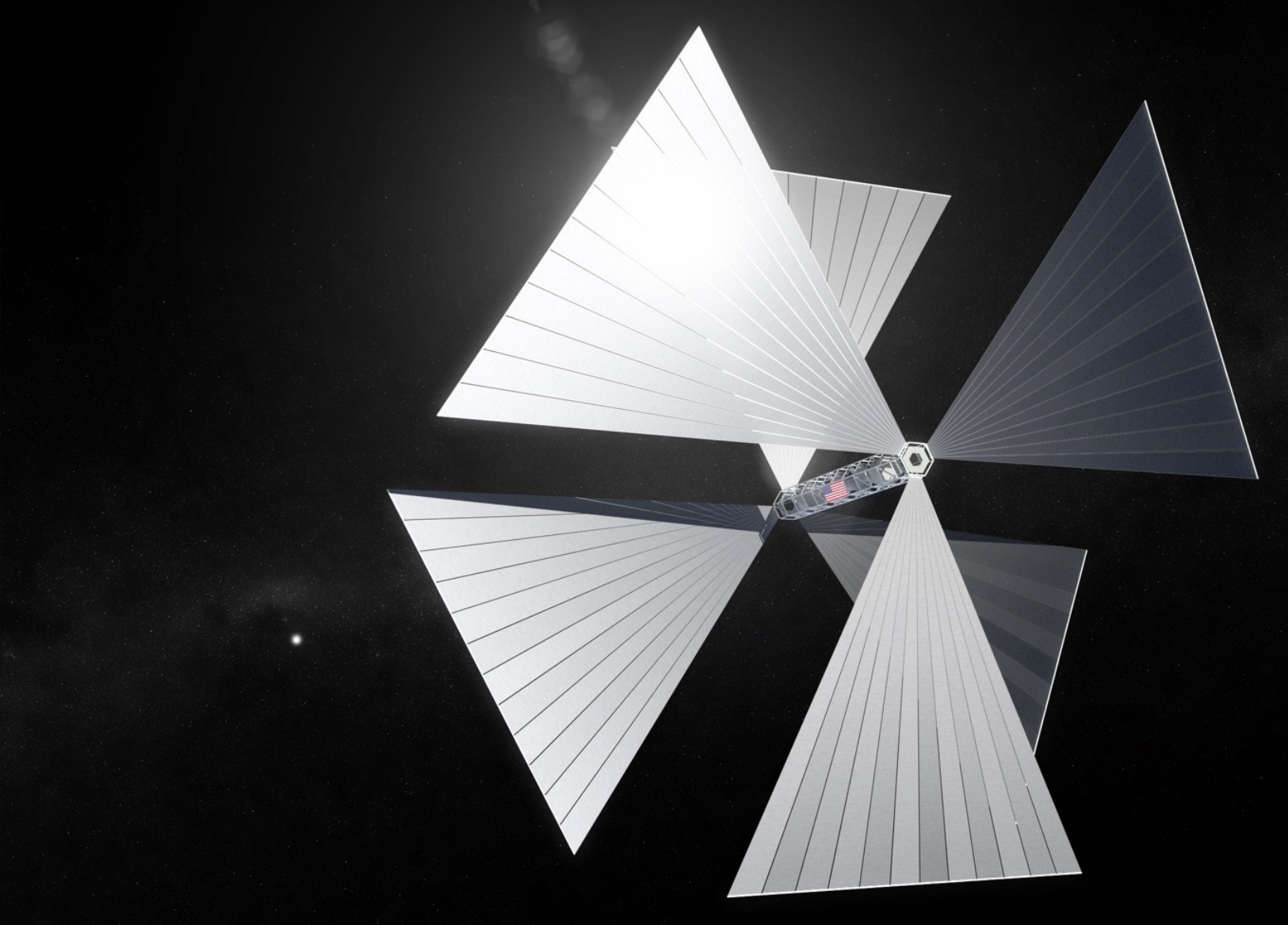}
	  \caption{Xplore's Lightcraft\textsuperscript{TM} advanced solar sail for rapid exploration of the solar system.}
	  \label{fig6}
	  \vskip 6pt 
%\end{figure}
%\begin{figure}%[<options>]
	\centering
	        \includegraphics[scale=0.70]{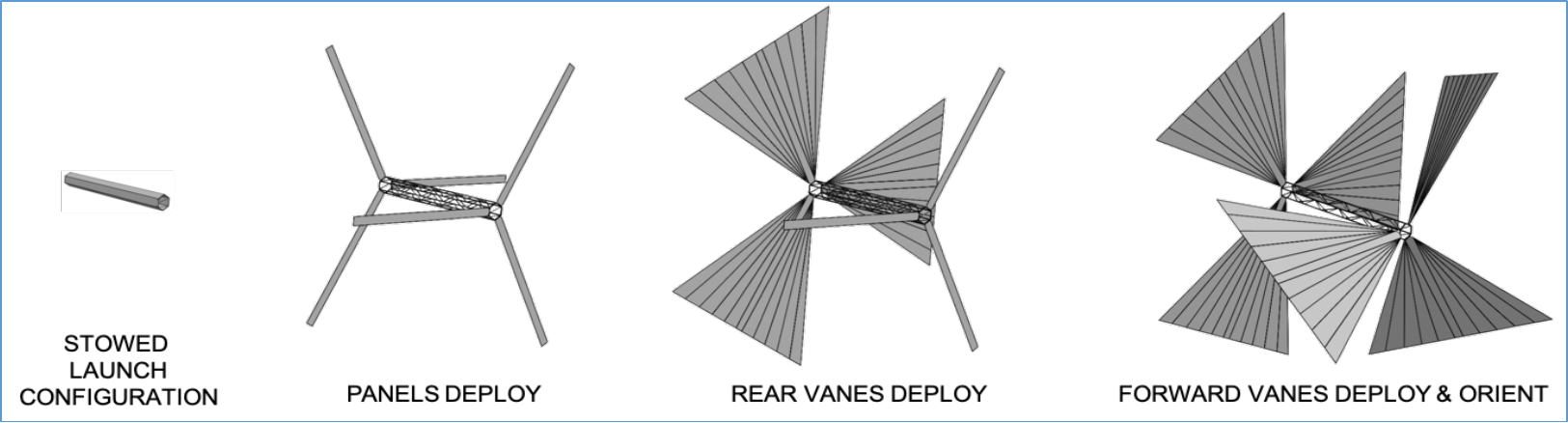}
	  \caption{Lightcraft\textsuperscript{TM} fractionated sail deployment and articulation sequence.}
	  \label{fig5}
	  \vskip 4pt 
\end{figure}

At JPL in the mid-1970s JPL has studied a sail that was 15 km in diameter (!) called a heliogyro. It was equivalent in size to a square sail that was 1/2 mile on each side. That was far too audacious.  The problem isn't just the size, it is whether the gossamer like structure can keeps it integrity and shape over years of flight, and whether it can be reasonably packaged and deployed however it is launched.  Based on experiences now building sailscraft and deploying and controlling things in space it seems that $10^4 ~{\rm m}^2$  may be as large as we should think about at present.  That would be a $100\times100$~${\rm m}^2$ sail, which actually might require boom stiffening or guy-wires.  A novel sailcraft design\footnote{A patented design invented by D. Garber of Xplore, Inc. and N. Barnes of L'Garde, Inc.} called the SunVane\textsuperscript{TM}  is now being studied which uses multiple sails (or vanes) that are smaller, easier to package and more controllable (Figs.~\ref{fig5}, \ref{fig6}). The SunVane advanced sailcraft design is being developed by Xplore Inc. as the Lightcraft\textsuperscript{TM} vehicle to enable rapid missions across cislunar space, to the Outer Planets and their moons and rendezvous with transient interstellar objects.
 
\subsection{How light can we make the spacecraft?}

 There are there parts to that -- how thin can the sail material be and how small can the spacecraft be and the relevant boom mass. While there are microsats less than 10 kg in mass, they are not yet robust enough for many years of interplanetary flight.  But even when they are, to work in the outer solar system and interstellar medium they will need to have a power source independent of the Sun -- today this is nuclear power.  Even advanced nuclear batteries, or using the decay heat of long half-life radioactive elements in a generator we will probably be limited to spacecraft that weigh tens of kilograms -- perhaps 50 kg, especially if we are carrying good science instrumentation.  The sail material might be as thin as 1 micron with density less than $1~{\rm g/m}^2$.  A sail of $10^4 ~{\rm m}^2$ would then weigh about 10 kg.  That still leaves 40 kg for the spacecraft, its sail structure and science instruments and power -- not yet today's technology  but probably possible in the not too distant future.  A $10^4$ square meter sail on a 50 kg spacecraft has an area, $A$, to mass, $m$, ratio of $A/m=200~{\rm  m}^2/{\rm kg}$.  
 
 \subsection{Can we get close to the Sun?}

This also will be determined by the sail and spacecraft separately.  For the latter, we will have to protect all the electronic components with shielding.  That will add too much mass, if we are too close.  

 \begin{figure}%[<options>]
	\centering
	        \includegraphics[scale=0.55]{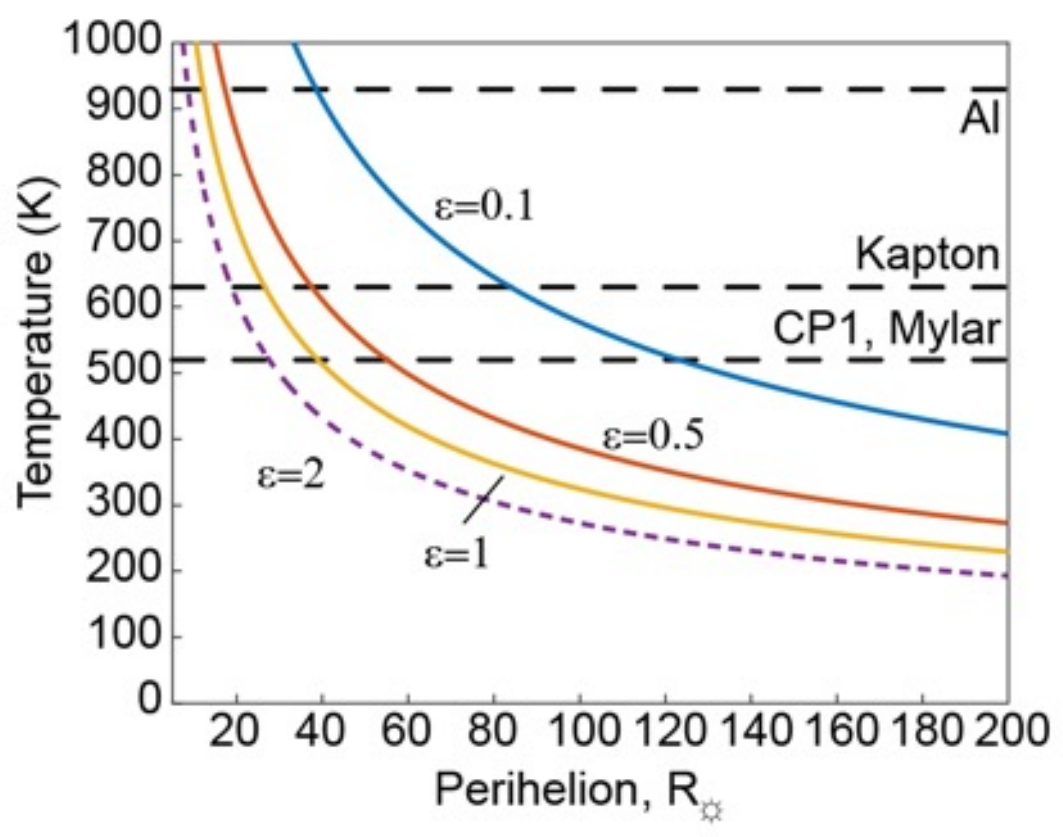}
	  \caption{Calculated temperature of the aluminized sail with perihelion distance for several different values of backside emissivity, $\epsilon$. Melting temperatures of typical sail polymer materials and aluminum are shown.}
	  \label{fig7}
\end{figure}

 \begin{figure}%[<options>]
	\centering
	        \includegraphics[scale=0.45]{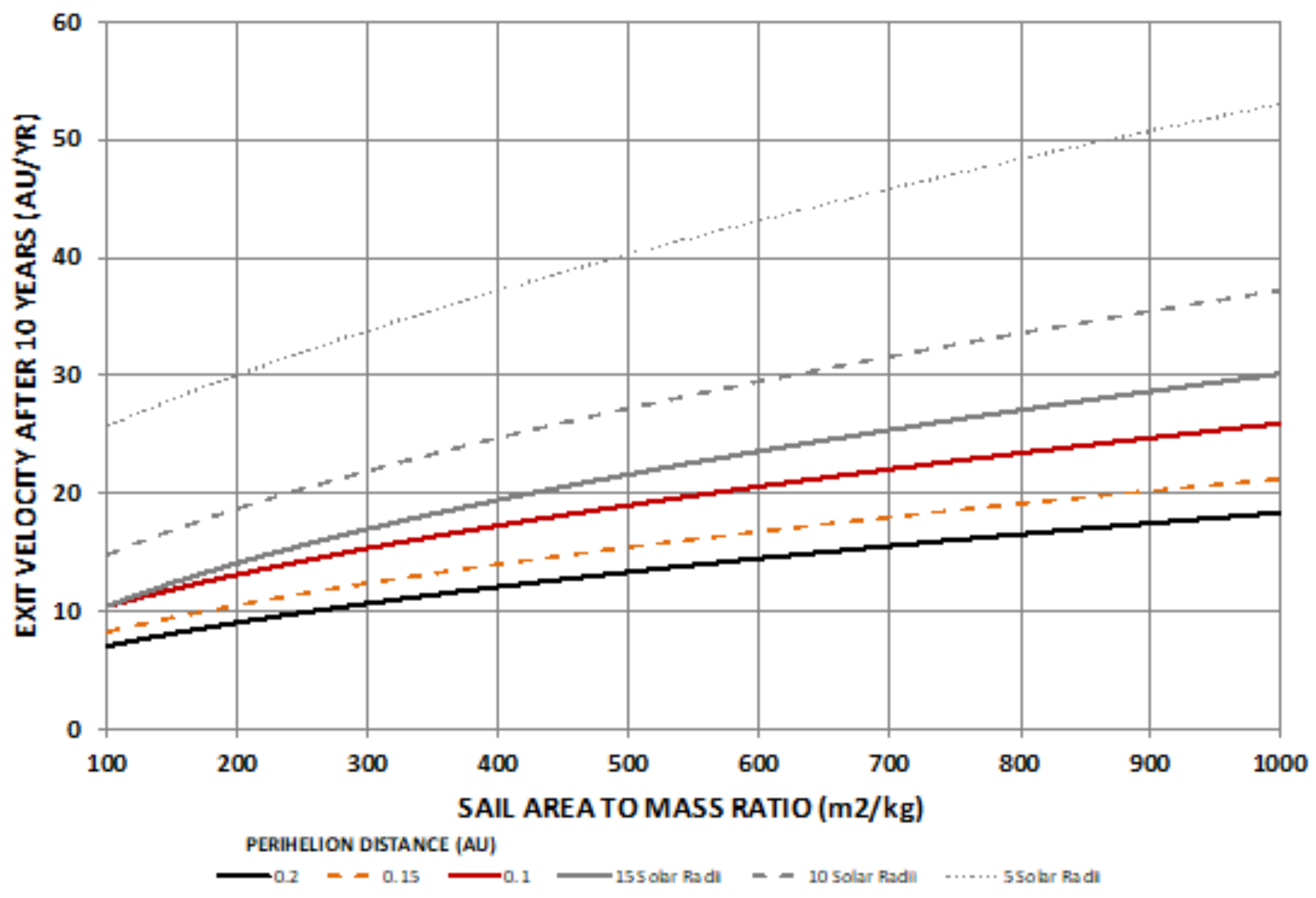}
	  \caption{Solar system exit velocity in AU/year vs. sail area to total spacecraft mass ratio for various perihelion distances of the trajectory.}
	  \label{fig8}
\end{figure}

The closest approach a sailcraft is capable of getting to the sun will also be determined separately by materials and architecture of the sail systems and by spacecraft design. In particular, the spacecraft needs to be designed to protect sensitive instruments and components from the damage caused by the sunlight (which already at 0.1 AU perihelion, that is $\sim20R_\odot$, reaches $100\times$ that of solar flux reaching the Earth). Here, concepts similar to James Webb Space Telescope's Sunshield\footnote{{https://jwst.nasa.gov/content/observatory/sunshield.html}, {https://www.vice.com/en/article/9akddp/this-sunshield-will-keep-the-most-powerful-space-telescope-ever-from-frying}} 
and Parker Solar Probe's Thermal Protection System\footnote{https://www.nasa.gov/sites/default/files/atoms/files/parkersolarprobe\_presskit\_august2018\_final.pdf} \cite{Ercol-Holtzman:2020}. Such a protection system against damaging solar radiation may be comprised of multilayer or composite materials \cite{Youngquist-Nurge:2018}. A careful attention in design of such a structure needs to be paid to its mass, which may take a significant fraction of the spacecraft. 

In addition to the protection of instruments, the sail systems too must withstand direct illumination by the Sun.  For the latter, we will have to protect all the electronic components with shielding.  That will add to much mass, if we are too close. Present day sail materials are made of aluminized polymer films (50--100~nm thick aluminum coated atop of a few micron thick polymer, such as CP1, Mylar, or Kapton \cite{Vulpetti-Johnson-Matloff:2015}). Aluminum on average absorbs 8--10$\%$ of sunlight \cite{Ercol-Holtzman:1984}. This fraction of absorbed sunlight is converted into heat, which, if not removed, will cause materials melting. In Fig.~\ref{fig7}, we plot anticipated temperature with perihelion distance for different values of backside emissivity. Present day sails have emissivity of $\epsilon \simeq 0.1$ which prevents from getting close to the sun. High melting temperature Kapton ($\sim670$~K) based sails can get to $\sim80R_\odot$ perihelion (i.e., $\sim0.37$ AU). With a proper thermal engineering which can increase the backside emissivity, such as an aluminized Kapton sail may be capable of getting to 30--40\,$R_\odot$. Getting even close to the sun would require replacing Kapton with a higher temperature inorganic material. In this case the sail damage is limited by the melting point of aluminum ($\sim930$~K). With proper thermal design such a sail is capable of getting to 12--15\,$R_\odot$. 

The sailcraft $v_{\rm inf}$ (i.e., solar system hyperbolic excess velocity), in AU/year, that can be achieved as a function of these three parameters is shown in Fig.~\ref{fig8}.  Current technology seems to allow reaching 20 AU/year,  about six times faster than the fastest escape velocity ever achieved the mission would take 30 years to reach the beginning of the solar gravity lens focal line and another fifteen or so years to provide enough integration time for continuous observations of the exoplanet while it orbits its parent star. That is a long-term science objective, although the payoff,  imaging the surface of an exoplanet at kilometer scale resolution, would be uniquely important.  

Advanced technology may permit sails the size of a football field and spacecraft the size of modern CubeSats, and coming close to the Sun with exotic materials of high reflectivity and able to withstand the very high temperatures.   That might permit going twice as fast, 40 AU/year or higher.  If we can do that it will be worth waiting for.   With long mission times, and with likely exoplanets in several different star systems being important targets of exploration we may want to develop a low cost, highly repeatable and flexible spacecraft architecture -- one that might permit a series of small missions rather than one with a traditional large, complex spacecraft. The velocity might also be boosted with a hybrid approach, adding an electric propulsion to the solar sail.  

As noted before, we need to carry a small amount of electric propulsion anyway to have maneuvering capability far from the sun near the solar gravity lens focus.   The approach also permits redundancy for increased reliability.  The multiple smallsat-sailcraft, close-to-the-Sun perihelion is a new mission design approach and depends on the viability of the interplanetary microsat, and on solar sailing.  

Another innovation is the multiple spacecraft approach.  Of course, all exploration involves multiple spacecraft -- just as ocean exploration involved multiple ships and voyages.  The exploration of our solar system has been carried out by many spacecraft -- dozens at the Moon, Mars, and Venus, and several at Mercury, Jupiter, Saturn with more planned.  Similarly multiple telescope missions observe the Universe.  But going to the far reaches of the solar system, and into the interstellar medium (ISM) takes much longer times.  

Observing exoplanets with the SGL not only takes a long time, but there are many different planets we might want to explore.  The small spacecraft approach allows that exploration to be carried out together, rather than sequentially waiting for one mission to be over before starting another -- like the Nina, Pinta and Santa Maria, the multiple ship approach provides increased reliability and robustness to exploration.  This new approach would permit the exploration of dozens of potentially habitable (or even inhabited) exoplanets in the second half of this century with high-resolution detailed observations and direct imaging.  

Right now, with today's technology, it would probably take several dozens of years to reach the solar gravity lens focus and get images of an exoplanet -- but with smaller spacecraft and advanced sail materials to go close to the Sun, we might be able to reach speeds greater than 40 AU/year and send multiple spacecraft to observe multiple exoplanets all for less than a cost of single space telescope (e.g., Hubble or James Webb telescopes).  We envision a series of small sailcraft to be Exoplanet Observers in the latter half of this century.  

 \section{Conclusions}
 \label{sec:concl}

The SGLF mission design calls for several innovations -- all within current technology, and more importantly already in development.  First, the interplanetary microsat sailcraft must be demonstrated to fly out of the solar system -- we call that the LightCraft Flight Test.  Next we would propose  to fly a science mission with capable instruments in the solar system to prove our concept of operations.  We choose a mission that is impossible to do (at reasonable cost) in any other way but is enabled by the microsat-solar sail:  a rendezvous with an ISO \cite{Garber-etal:2021-ISO-paper}.  Then we would be ready to fly out of the solar system to solar gravity lens focal line, passing by KBOs and the Heilopause on the way \cite{Turyshev-etal-BT:2020}.  

The key innovative ideas:
\begin{itemize}
\item Very lightweight interplanetary smallsats;
\item A new multi-sail design for solar sailcraft;
\item The deconvolution of images created by pixel-by-pixels sampling the image produced by the SGL over 10s of years and 100s of AU;
\item Multiple spacecraft to be assembled or networked into a mission-capable spacecraft with large enough telescope, power and microthruster propellant to carry out imaging operations, gather and compress the data, and communicate it back to Earth some 100 billion km distant.  
\end{itemize}

 An incremental approach to prove each of the innovations is necessary, each one increasing mission capability and reliability.  The incremental approach dovetails well with a step-by-step approach of deeper space missions of more and more complexity.  Each of the steps can yield technical and scientific benefits on the way to the grand goal of imaging a putatively habitable world.  
 
The first step, as mentioned above, is proving the concept of achieving record-breaking speed out of the solar system with interplanetary microsats and using the solar sail to fly close to the Sun and provide the necessary velocity.  It is a technology demonstration -- no ambitious science instruments for the payload.  In fact, this could be a cultural mission of popular interest -- science, art and public engagement connected to the interstellar goal and sending the spacecraft fast out of the solar system.  Such a mission might carry music and art, messages and recordings, DNA and/or other microbial information within a sub-kilogram payload on a spacecraft headed out of the solar system. We are studying the formation of a public-private partnership combining NASA interest in the technology and science missions with public interest in the exit of the solar system.   The first technology test of the basic mission design to exercise the microsat technologies, solar sail control and communications and telemetry links with Earth.  Small cis-lunar missions could also test the spacecraft subsystems and multi-spacecraft operations in parallel.  Succeeding precursor steps will add scientific payload and test other technologies for advanced exploration hundreds of AU away.  

Two scientific objectives that particularly need the capabilities of fast flight, multi-spacecraft,  and microsats are ISOs and KBOs.  Both objectives are important in the field of planetary science, and both could serve as excellent interstellar precursor missions.  We have mentioned ISOs, when discussing the first one discovered, the strange looking Oumuamua.  It's discovery in 2017 was followed another one just two years later, Borisov in 2019.  It was more conventional, looking and behaving like a comet from interstellar space (as its discovery article was titled in the scientific literature).  ISOs can't be predicted (at least not far in advance) because they are very small (like comets and asteroids) and very fast (flying on hyperbolic trajectories a single time through the solar system.  The two thus far discovered ISOs, Oumuamua reached a perihelion of 0.25 AU, with an $v_{\rm inf}\sim 5$ AU/year   while Borisov reached only about 2 AU (further out that Mars) with a $v_{\rm inf}\sim 6$ AU/year.  We can guess that many more ISOs will be discovered in the future, astronomer say perhaps about one per year.  Getting up close and to explore an ISO would be interesting and important -- a close up look of something from another star system.  If we could rendezvous and land on one, we could bring back samples to analyze in Earth-based laboratories, or we could ``ride'' on it to interstellar space. But a mission to an ISO requires (i) extremely fast reaction time to a discovery and (ii) a spacecraft capable of going faster than the ISO.  These can be done with the mission design described herein \cite{Garber-etal:2021-ISO-paper}.  

Getting to and operating in the Kuiper Belt to study several KBOs would take us almost into the interstellar medium.  (The Kuiper Belt extends from Pluto at about 40 AU to perhaps 120 AU, whereas the interstellar medium begins around 140 AU).   Distances out there are immense, and a survey and study of KBOs would require many spacecraft.  This is another problem that the smallsat-solar sail design can solve -- the small spacecraft will be relatively inexpensive and multiple launches from the Earth would not require big and expensive launch vehicles.   While the flybys of KBOs would be quick (e.g., 25 km/sec, passing by them in just minutes, discoveries could be followed up with other reconnaissance spacecraft and the comparative data from many KBO flybys would be very valuable.   The ISO and KBO interstellar precursors could be carried out in the late 2020s and early 2030s, enabling technology readiness and development of the SGLF mission and Exoplanet Observers to be done by the middle 2030s, with the potential of obtaining images of habitable or inhabited exoplanets by the 2060s and 2070s.  A possible sequence of technology used in such missions would then retire the technology risk for the Exoplanet Observers. To summarize the steps:

 \begin{figure}%[<options>]
	\centering
	        \includegraphics[scale=1.55]{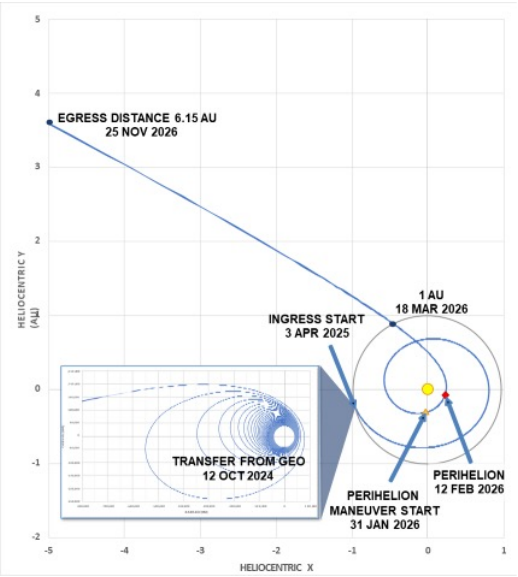}
	  \caption{The trajectory for the Lightcraft Demonstration Flight.}
	  \label{fig9}
	  \vskip 2pt 
\end{figure}

\begin{figure}%[<options>]
	\centering
	        \includegraphics[scale=0.70]{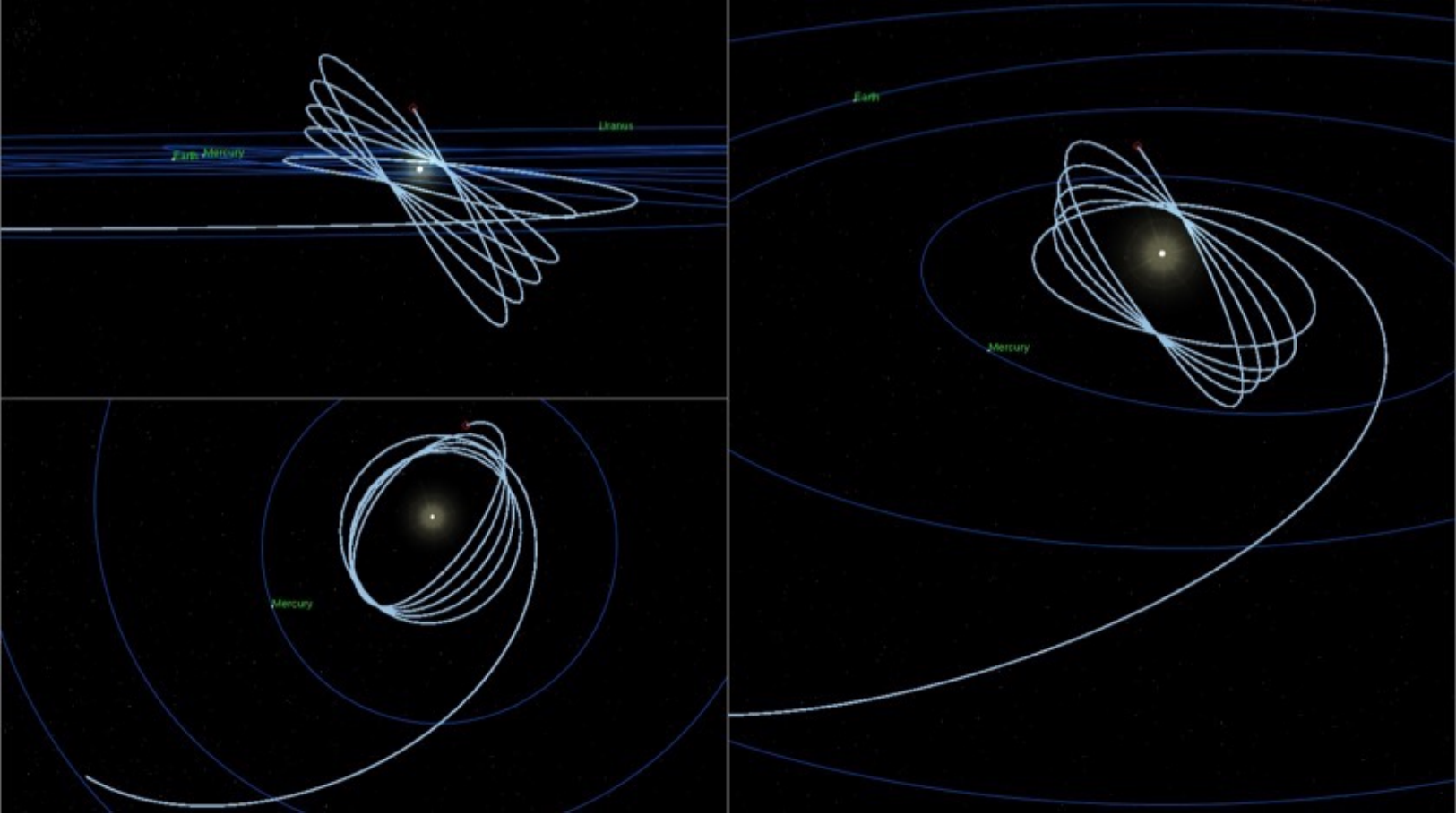}
	  \caption{Views of the heliocentric trajectory for a mission to intercept an Interstellar Object.}
	  \label{fig10}
\end{figure}

\begin{itemize}
\item  A Lightcraft\textsuperscript{TM} Demonstration Mission to exercise and prove the technology of an interplanetary microsat sailcraft to exit the solar system faster than any previous vehicle, with goal of 5--6 AU/year, see Fig.~\ref{fig9}.  The mission would carry only popular interest payloads and be conducted as a public-private partnership, with NASA support for the technology demonstration and private support for the popular interest. Total mass of sailcraft $\approx 6$ kg \cite{Garber-etal:2021-ISO-paper}. 
\item  A solar system mission, possibly to rendezvous with a newly discovered interstellar object to another small body in or beyond the asteroid belt.  Velocity goal of $>7$~AU/year, with science payload $<10$~kg, see Fig.~\ref{fig10}.
\item Exoplanet Observers -- mission capability to reach the focal line of the solar gravity lens in $<25$ years. Exoplanet imaging with the SGL using 1--2~m coronagraph/telescope, optical communications, small radioisotope power, electric micro-thrusters \cite{Garber-etal:2021-ISO-paper}.
\end{itemize}

%By this point the mission design would also enable a Comet Halley to rendezvous during its 2061 apparition. Whether or not interstellar travel remains a bridge too far, the ability to explore many habitable and interesting exoplanets in the latter quarter of this century will certainly open up a new age of science and exploration, and a new understanding about the possibilities of extraterrestrial life.  

%\section{Conclusions}

A new mission design to permit fast solar system exit missions and enable high-resolution, multi-pixel imaging of many potentially habitable exoplanets has been described.  It relies on the new technologies of interplanetary microsats and solar sails.  A new sail vehicle design is also proposed.  A near-term, low-cost public-private partnership for a demonstration mission would prove the mission design concept and engage the public with a fast exit of the solar system.  Other missions enabled include a rendezvous with an ISO and flybys of KBOs.  The ultimate payoff is that which drives space exploration, seeing life on another world.  

\begin{acknowledgements}
The work described here, in part, was carried out at the Jet Propulsion Laboratory, California Institute of Technology, under a contract with the National Aeronautics and Space Administration. Additional funding has been provided by the Aerospace Corporation.  The NIAC Phase III study from which this work was derived also included Breakthrough Initiatives, Xplore Inc., and Cornell Tech (a campus of Cornell University). We thank the NASA Innovative Advanced Concepts (NIAC) program for their support.  
%\textcopyright 2021. All rights reserved. Pre-decisional information -- for planning and discussion purposes only.

\end{acknowledgements}


\bibliographystyle{spphys}       % APS-like style for physics

%\bibliography{mission-design} 

\end{document}